\documentclass{jps-cp}
\usepackage{txfonts} %Please comment out this line unless the txfonts package is availabe in your LaTeX system.
\newcommand\newblock{ }
\usepackage{braket}
\usepackage{bm}
\usepackage{color}

\title{Study of hadron interactions and compositeness}

\author{Tetsuo \textsc{Hyodo}$^{1,2}$}

\inst{$^{1}$Department of Physics, Tokyo Metropolitan University, Hachioji 192-0397, Japan \\
$^{2}$RIKEN Interdisciplinary Theoretical and Mathematical Science Program (iTHEMS), Wako 351-0198, Japan}

\email{hyodo@tmu.ac.jp}

\recdate{February 21, 2019}

\abst{Plenty of hadrons have been established experimentally, yet the nonperturbative nature of the strong interaction complicates a comprehensive understanding of their internal structure, particularly for exotic hadrons that extend beyond conventional mesons and baryons. One prominent candidate for the internal structure of exotic hadrons is the hadronic molecule, a loosely bound system of hadrons analogous to atomic nuclei. Understanding such systems requires precise knowledge of hadron interactions, which traditional scattering experiments struggle to provide, especially in the low-energy region. Recent advances, including precise baryon-baryon interaction measurements, femtoscopy techniques that probe momentum correlations between particles, and first-principles lattice QCD calculations, have significantly improved our understanding of hadron interactions. Here, we review these recent developments, demonstrate the successful application of femtoscopy to antikaon-nucleon interactions, utilize the concept of compositeness to quantify the hadronic molecular component of the $\Lambda(1405)$, and discuss both experimental and theoretical prospects, including future studies at J-PARC.}

\kword{exotic hadrons, hadron interactions, femtoscopy, compositeness, $\Lambda(1405)$}

\begin{document}
\maketitle

%%%%%%%%%%%%%%%%%%%%
\section{Introduction}
%%%%%%%%%%%%%%%%%%%%

% exotic hadrons
Currently, more than 400 species of hadrons have been experimentally observed~\cite{ParticleDataGroup:2024cfk}. The number of observed hadrons continues to grow year by year, with over 20 new hadrons reported within the last two years. These hadrons are thought to emerge from the strong interaction between quarks and gluons. However, a comprehensive understanding of how hadrons are constructed from quarks and gluons has not yet been fully achieved due to the nonperturbative nature of the strong interaction. In particular, recent experiments suggest the existence of exotic hadrons~\cite{Hosaka:2016pey,ExHIC:2017smd,Guo:2017jvc,Brambilla:2019esw,Hyodo:2020czb}, which are considered to possess internal structures that go beyond those of conventional mesons and baryons.

% hadronic molecules
Various internal structures have been proposed for exotic hadrons, including multiquark states, hadronic molecules, and gluon hybrids. Among them, hadronic molecules are of particular interest. These are loosely bound systems of hadrons, formed through hadron-hadron interactions. It is well known that atomic nuclei are self-bound systems of nucleons, held together by the nuclear force, rather than being directly composed of quarks and gluons. In an analogous manner, the binding mechanism of hadronic molecules is driven by interactions between hadrons. Consequently, understanding hadron interactions has become an increasingly important topic, as it provides key insights into the nature of hadronic molecules.

% hadron interactions
Traditionally, hadron interactions have been studied through scattering experiments. However, these experiments are limited to only a small subset of hadron pairs, and the statistical precision is often poor, especially in the low-energy region. Remarkably, several new techniques have recently been developed to probe hadron interactions. The J-PARC E40 collaboration has performed precise measurements of baryon-baryon interactions in the strangeness sector~\cite{J-PARCE40:2021bgw,J-PARCE40:2021qxa,J-PARCE40:2022nvq}. Another powerful approach is the study of momentum correlations between particle pairs emitted in high-energy collisions, known as the femtoscopy technique, which provides access to hadron interactions that are difficult to explore with conventional scattering experiments~\cite{ExHIC:2017smd,Fabbietti:2020bfg}. On the theoretical side, first-principles calculations using lattice QCD also provide crucial insights into exotic hadrons and hadron interactions~\cite{Ishii:2006ec,Iritani:2018sra,Lyu:2023xro}. These new approaches are shedding fresh light on hadron interaction studies.

% compositeness
In characterizing the internal structure of hadrons, a key concept is the compositeness, which quantifies the fraction of the hadronic molecular component. The concept was originally introduced in the 1960s to distinguish between elementary particles and composite particles~\cite{Weinberg:1965zz}. In recent years, it has attracted renewed attention for its applicability to the study of exotic hadrons~\cite{Baru:2003qq,Hyodo:2013nka,Oller:2017alp,vanKolck:2022lqz,Kinugawa:2024crb}. In particular, it is known that the compositeness of a near-threshold $s$-wave state can be determined from the scattering length and the eigenenergy via the weak-binding relation~\cite{Weinberg:1965zz,Kamiya:2015aea,Kamiya:2016oao,Kinugawa:2022fzn}. This emphasizes the importance of determining quantities such as the scattering length and eigenenergy through hadron scattering analyses.

% in this contribution
In this contribution, we review the current status of hadron interaction studies, with a focus on recent developments. In particular, we highlight the $\Lambda(1405)$ as a key example of an exotic hadron that is intimately related to the antikaon-nucleon interaction. We present the successful application of the femtoscopy technique to this interaction, achieved through close collaboration between theory and experiment~\cite{ALICE:2019gcn,Kamiya:2019uiw,ALICE:2021szj,ALICE:2022yyh}.
Furthermore, we discuss methods to determine the $\bar{K}N$ compositeness of the $\Lambda(1405)$ from experimental observables~\cite{Kamiya:2015aea,Kamiya:2016oao}. Finally, we outline future prospects for hadron interaction studies, including upcoming experiments at J-PARC.

%%%%%%%%%%%%%%%%%%%%
\section{Exotic hadrons and $\Lambda(1405)$}
%%%%%%%%%%%%%%%%%%%%

% hadrons
Hadrons have historically been classified into two categories: mesons, which consist of a quark-antiquark pair ($\bar{q}q$), and baryons, which consist of three quarks ($qqq$). For example, the proton ($p$) and neutron ($n$), which form atomic nuclei, and the charged pion ($\pi^{+}$), which mediates the nuclear force, are composed of $u$ and $d$ quarks as:
\begin{align*} 
   p& \sim uud,\quad n \sim udd,\quad \pi^{+} \sim \bar{d}u .
\end{align*} 
This classification has provided a successful framework for describing the hadron spectrum.

% exotic hadrons
However, with recent advances in experimental techniques, exotic hadrons, which do not fit into this conventional classification scheme, have been discovered and are attracting considerable attention. Experimental evidence for the existence of such new particles has been reported one after another, and various theoretical ideas have been proposed to interpret these findings, leading to active research efforts in this field~\cite{Hosaka:2016pey,ExHIC:2017smd,Guo:2017jvc,Brambilla:2019esw,Hyodo:2020czb}. The discovery of exotic hadrons indicates that hadrons can possess more diverse internal structures than previously expected. Elucidating their internal structures is thus expected to play a crucial role in deepening our understanding of low-energy QCD.

% Tcc
A striking example of an exotic hadron is $T_{cc}(3875)^{+}$, reported by the LHCb collaboration in 2021~\cite{LHCb:2021vvq,LHCb:2021auc}. $T_{cc}(3875)^{+}$ has been identified as a peak in the invariant mass distribution of the $D^{0}D^{0}\pi^{+}$ system in high-energy proton-proton collision experiments. Its decay into $D^{0}(\sim\bar{u}c) + D^{0}(\sim\bar{u}c) + \pi^{+}(\sim\bar{d}u)$ implies the minimal quark content:
\begin{align*} 
   T_{cc}(3875)^{+}& \sim cc\bar{u}\bar{d}.
\end{align*} 
Because the quark-antiquark pairs in the $cc\bar{u}\bar{d}$ cannot be annihilated further, this state is clearly distinct from ordinary hadrons, and is classified as a doubly charmed tetraquark state.

\begin{figure}[tb]
\centering
\includegraphics[width=5cm]{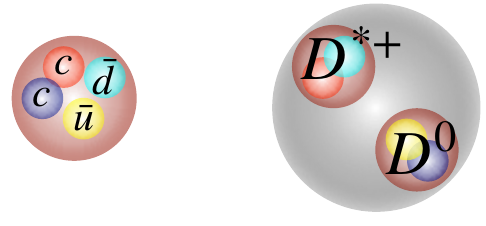}
\caption{Schematic illustration of the multiquark picture (left) and the hadronic molecule picture (right) for $T_{cc}(3875)^{+}$.}
\label{fig:schematic}
\end{figure}

% internal structure
Two of the most widely discussed internal structures for exotic hadrons are multiquark states and hadronic molecular states. The multiquark state is regarded as a single-hadron state where all quarks form a tightly bound system. In contrast, hadronic molecular states are loosely bound systems composed of multiple hadrons, and are thus considered composite (two-hadron) states. For instance, in the multiquark picture, $T_{cc}(3875)^{+}$ is described as a tightly bound $cc\bar{u}\bar{d}$ system (Fig.~\ref{fig:schematic}, left). On the other hand, in the hadronic molecule picture, a $c\bar{d}$ pair forms a $D^{*+}$, and a $c\bar{u}$ pair forms a $D^{0}$, with $T_{cc}(3875)^{+}$ emerging as a bound state due to the interaction between $D^{*+}$ and $D^{0}$ (Fig.~\ref{fig:schematic}, right).

% mixing
To proceed with more quantitative discussions, it is necessary to consider the mixing of different configurations. A physical state is generally a superposition of components with the same quantum numbers, meaning that multiquark and hadronic molecular components do not appear as mutually exclusive alternatives. For example, the physical $T_{cc}(3875)^{+}$ state can be described as
\begin{align*}
   \ket{T_{cc}(3875)^{+}}=c_{1}\ket{cc\bar{u}\bar{d}}
   +c_{2}\ket{D^{0}D^{*+}}+\dotsb ,
\end{align*}
which is a superposition of the multiquark and hadronic molecular components.

% Lambda(1405)
When considering the mixing of states, even those that were initially thought to be described by three quarks or a quark-antiquark pair can acquire structures similar to multiquark states or hadronic molecules by incorporating additional $\bar{q}q$ pairs. In fact, among hadrons that have been known for a long time, there are also candidates for exotic hadrons. A well-known example is the $\Lambda(1405)$, a negative-parity baryon containing a strange quark~\cite{Hyodo:2011ur,Meissner:2020khl,Mai:2020ltx,Hyodo:2020czb,Sadasivan:2022srs}. The minimal quark content of the $\Lambda(1405)$ is $uds$, yet there are several reasons to consider that the structure of the $\Lambda(1405)$ is predominantly a $\bar{K}N$ molecular state. First, in constituent quark models, which systematically describe the excitation spectrum of conventional baryons as three-quark states, the predicted mass of the $\Lambda(1405)$ significantly deviates from the experimental result, suggesting that the $\Lambda(1405)$ has a structure different from ordinary three-quark states. Second, when considered relative to the $\bar{K}N$ threshold, the $\Lambda(1405)$ can be regarded as a bound state with a binding energy of approximately 15 MeV~\cite{ParticleDataGroup:2024cfk}. This supports the interpretation that the $\Lambda(1405)$ is a quasibound state formed by the attractive interaction between $\bar{K}$ and $N$. In fact, the $\Lambda(1405)$ was originally predicted based on theoretical analyses of the $\bar{K}N$ scattering data~\cite{Dalitz:1959dn}.

% chiral coupled-channel approach
In modern approaches, the $\Lambda(1405)$ is described as a resonance state arising in the $\bar{K}N$ scattering within the chiral coupled-channel framework, which combines chiral perturbation theory and scattering theory~\cite{Kaiser:1995eg,Oset:1997it,Oller:2000fj,Hyodo:2011ur,Meissner:2020khl,Mai:2020ltx,Hyodo:2020czb,Sadasivan:2022srs}. In particular, since the 2010s, next-to-leading order (NLO) analyses have been performed, incorporating precise experimental data from kaonic hydrogen measurements, leading to a highly accurate determination of the eigenenergy of the $\Lambda(1405)$~\cite{Ikeda:2011pi,Ikeda:2012au,Mai:2012dt,Guo:2012vv,Mai:2014xna,Feijoo:2018den}. More recently, the next-to-next-to-leading order (NNLO) analysis have also been conducted~\cite{Lu:2022hwm}. In the latter half of this contribution, we review recent studies on the $\bar{K}N$ interaction and the internal structure of the $\Lambda(1405)$.

%%%%%%%%%%%%%%%%%%%%
\section{Study of hadron interactions}
%%%%%%%%%%%%%%%%%%%%

% traditional scattering experimental
The traditional study of hadron interactions originates from the investigation of the nuclear force through the $NN$ scattering. In nuclear force studies, the $NN$ scattering experiments are performed to measure the differential cross sections, from which phase shifts are extracted via partial wave analysis, and theoretical models are constructed based on these results. Applying this scattering experiment approach to the study of exotic hadrons and hadronic molecules faces several difficulties. First, scattering experiments require stable targets and well-controlled beams, limiting the systems that can be studied to cases such as $\pi N$, $KN$, and $\bar{K}N$ scatterings. In particular, scattering experiments in sectors involving charm quarks, which are essential for the study of exotic hadrons, are practically impossible. Moreover, at low energies, the instability of beam particles results in limited experimental data with low statistics. For instance, the low-energy $\bar{K}N$ scattering data, which are closely related to the $\Lambda(1405)$, are also statistically insufficient to accurately constrain the $\bar{K}N$ interactions.

% new developments
It is remarkable that several new techniques have recently been developed to study hadron interactions. In particular, significant progress has been made for
\begin{itemize}
\item high-precesion experiments of the baryon-baryon scattering with strangeness,
\item femtoscopy technique for the two-particle momentum correlation functions, and 
\item lattice QCD calculations for hadron scatterings.
\end{itemize}
In the following, we introduce these topics.

% J-PARC E40
The J-PARC E40 collaboration has performed precise measurements of baryon-baryon interactions in the strangeness sector~\cite{J-PARCE40:2021bgw,J-PARCE40:2021qxa,J-PARCE40:2022nvq} using the high-intensity $\pi^-$ beam from the K1.8 beamline. By producing $\Sigma^{+}$ and $\Sigma^{-}$ hyperons via the $\pi^- p \to K^+ \Sigma^\pm$ reactions and detecting their scatterings with protons in a liquid hydrogen target, differential cross sections of $\Sigma^{-}p\to \Lambda n$, 
$\Sigma^{-}p\to \Sigma^{-}p$, and 
$\Sigma^{+}p\to \Sigma^{+}p$ are measured. Thanks to the high-intensity beam and 
detector system, the experiment has achieved the results with significantly higher statistics compared to previous measurements. These results provide crucial constraints on the baryon-baryon interaction in the strangeness $S=-1$ sector. In fact, the results of the J-PARC E40 are implemented in the latest next-to-next-to leading order chiral perturbation theory for the hyperon-nucleon interaction to determine the low-energy constants~\cite{Haidenbauer:2023qhf}.

%\begin{figure}[tb]
%\centering
%\includegraphics[width=5cm]{schematic.pdf}
%\caption{Schematic illustration of the multiquark picture (left) and the hadronic molecule picture (right) for $T_{cc}(3875)^{+}$.}
%\label{fig:schematic}
%\end{figure}

% femtoscopy
The study of momentum correlations between particle pairs emitted in high-energy collisions, known as the femtoscopy technique, provides information on hadron interactions that are not accessible through traditional experiment. In the high-energy collisions where a large number of hadrons are produced, the momentum correlation of a given hadron pair reflects the final state interaction between the pair~\cite{ExHIC:2017smd,Fabbietti:2020bfg}. Theoretically, the momentum correlation function $C(\bm{q})$ is calculated using the Koonin-Pratt formula~\cite{Koonin:1977fh,Pratt:1984su,Pratt:1986cc,Bauer:1992ffu,Murase:2024ssm}:
\begin{equation}
    C(\bm{q}) = \int d^3r\ S(\bm{r})\ | \Psi^{(-)}_{\bm{q}}(\bm{r}) |^2,
    \label{eq:KPformula}
\end{equation}
where $S(\bm{r})$ is the source function characterizing the size and shape of the hadron emission source created by the collisions, and $\Psi^{(-)}_{\bm{q}}(\bm{r})$ is the relative wave function reflecting the final-state interactions. Therefore, by measuring the correlation function $C(\bm{q})$ of hadron pairs whose interactions are already known from scattering experiments, information on the source function $S(\bm{r})$ can be extracted. This was the original concept of femtoscopy~\cite{Lisa:2005dd}. In recent years, however, the method has evolved in the opposite direction; utilizing the accumulated knowledge of source functions to extract information on the interactions themselves from the measured correlation functions of various hadron pairs. In fact, the ALICE collaboration has measured the $p\Omega$ correlation functions with strangeness $S=-3$~\cite{ALICE:2020mfd}, and the $D^{-}p$, $D\pi$ and $DK$ correlation functions in the charm hadron sector~\cite{ALICE:2022enj,ALICE:2024bhk}. These channels are almost impossible to access in the usual scattering experiments. In addition, the femtoscopy technique allows us to determine the hadron interactions with unprecedented precision, as represented by the measurement of the $K^{-}p$ interactions. This will be the topic of the next section. 

% lattice QCD
On the theoretical side, first-principles calculations using lattice QCD provide important insights into hadron interactions. For instance, the HAL QCD method has been developed to calculate the potentials of hadron-hadron interactions on the lattice~\cite{Ishii:2006ec}. This method is applied to the $N\Omega$ system~\cite{Iritani:2018sra} and the $DD^{*}$ system~\cite{Lyu:2023xro} by the HAL QCD collaboration, successfully extracting the corresponding potentials. At the same time, the Baryon Scattering (BaSc) collaboration has performed coupled-channel $\pi\Sigma$-$\bar{K}N$ scattering calculations on the lattice~\cite{BaryonScatteringBaSc:2023zvt,BaryonScatteringBaSc:2023ori}, indicating the existence of two eigenstates in the energy region of the $\Lambda(1405)$. These studies demonstrate that lattice QCD techniques serve as a powerful tool for investigating hadron interactions.

%%%%%%%%%%%%%%%%%%%%
\section{$K^{-}p$ correlation functions }
%%%%%%%%%%%%%%%%%%%%

\begin{figure}[bt]
\centering
\includegraphics[width=8cm]{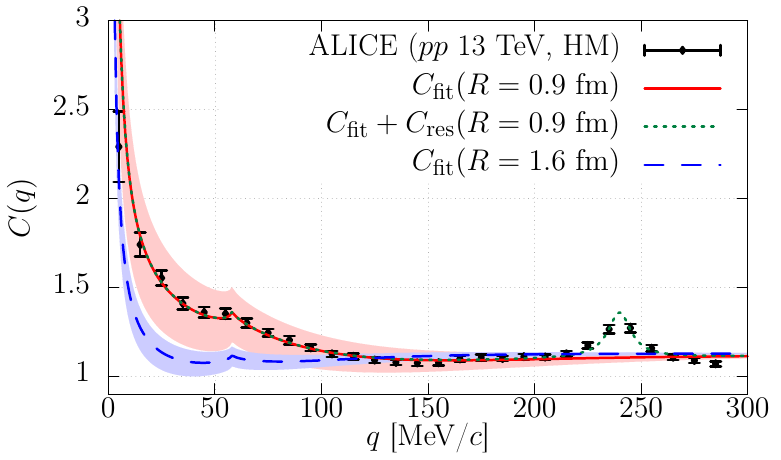}
\caption{Correlation function with the best fit values at $R=0.9 \ \mathrm{fm}$ (solid line). The fitting result with the $\Lambda(1520)$ contribution is shown by the dotted line. The dashed line shows the prediction with $R=1.6 \ \mathrm{fm}$. Its shaded area shows the uncertainty with respect to the variation of $\omega_{\pi\Sigma}$. For comparison, we also plot the corresponding area for the case with $R=0.9 \ \mathrm{fm}$.  The ALICE data set is taken from Ref.~\cite{ALICE:2019gcn}. Figure adapted from Ref.~\cite{Kamiya:2019uiw}.}
\label{fig:fit_predict}
\end{figure}

% accurate femtoscopy data
Let us now focus on a specific system, the $\bar{K}N$ interaction, which plays a crucial role in understanding the nature of the $\Lambda(1405)$ resonance. Traditionally, theoretical models for the $\bar{K}N$ interaction have been constrained by the $K^-p$ total cross section data obtained from old bubble chamber experiments. However, the quality of the low-energy data is limited, with relatively poor statistics. One of the key applications of femtoscopy is to provide new insights into the $\bar{K}N$ interaction. In fact, the ALICE collaboration has measured the $K^{-}p$ correlation function in high-energy proton-proton ($pp$) collisions~\cite{ALICE:2019gcn}. The precision of the data is remarkable, resolving the cusp at the $\bar{K}^{0}n$ threshold ($q\sim 58$ MeV$/c$) in the correlation function as shown in Fig.~\ref{fig:fit_predict}. Furthermore, data points are available even in the extremely low-energy region below the $\bar{K}^{0}n$ threshold. These high-quality correlation function data provide important new constraints on the $\bar{K}N$ interaction and the properties of the $\Lambda(1405)$.

% theoretical framework
To theoretically calculate the $K^-p$ correlation functions, three important extensions must be incorporated. First, the effect of channel coupling arises due to the transition from the $K^-p$ channel to the $\bar{K}^{0}n$, $\pi^{+}\Sigma^{-}$, $\pi^{0}\Sigma^{0}$, $\pi^{-}\Sigma^{+}$, and $\pi^{0}\Lambda$ channels. The coupled-channel generalization of the Koonin-Pratt formula has been formulated in Refs.~\cite{Lednicky:1997qr,Haidenbauer:2018jvl,Kamiya:2019uiw} and is expressed as
\begin{equation}
    C_{K^{-}p}(\bm{q}) = \int d^3r\ S_{K^{-}p}(\bm{r})\ | \Psi^{(-)}_{K^{-}p,\bm{q}}(\bm{r}) |^2
    +\sum_{i\neq K^{-}p}
    \omega_{i}\int d^3r\ S_{i}(\bm{r})\ | \Psi^{(-)}_{i,\bm{q}}(\bm{r}) |^2,
\end{equation}
where $S_i(\bm{r})$ is the source function for channel $i$, and $\omega_i$ represents the weight factor of channel $i$ relative to the $K^{-}p$ channel. Second, since the $K^-p$ channel is subject to both strong and Coulomb interactions, the effect of the Coulomb interaction must also be incorporated as a final-state interaction. Third, the energy difference between the $K^-p$ and $\bar{K}^0n$ thresholds due to the isospin symmetry breaking must be properly taken into account. While the effect of isospin symmetry breaking is typically neglected in calculations of strong interactions as it provides only a minor correction, the ALICE data are precise enough to resolve the $\bar{K}^0n$ cusp. Thus, to ensure consistency with the experimental precision, theoretical calculations must also include the effect of the isospin symmetry breaking.

% comparison with pp data
In Ref.~\cite{Kamiya:2019uiw}, theoretical calculations of the $K^{-}p$ correlation functions incorporating all three aforementioned extensions were performed and compared with the ALICE data (Fig.~\ref{fig:fit_predict}, solid line). The Kyoto $\bar{K}N$-$\pi\Sigma$-$\pi\Lambda$ potential~\cite{Miyahara:2018onh}, which reproduces the scattering amplitude of the chiral coupled-channel approach~\cite{Ikeda:2011pi,Ikeda:2012au}, is employed to calculate the scattering wavefunction $\Psi^{(-)}_{i,\bm{q}}(\bm{r})$. By appropriately choosing the source size $R\sim 1$ fm of the Gaussian source function and the $\pi\Sigma$ weight $\omega_{\pi\Sigma}\sim 2$, the experimental data of the ALICE correlation function was successfully reproduced with high precision. The obtained source size is consistent with the analysis of the $K^{+}p$ correlation, whose interaction is relatively well understood and is measured simultaneously by ALICE, while the value of $\omega_{\pi\Sigma}$ is in agreement with a simple statistical model estimation. This demonstrates that by employing a realistic $\bar{K}N$ potential and a properly formulated channel-coupled correlation function, the ALICE experimental data can be well reproduced.

% Detailed study of the source size dependence
The source size dependence of the correlation function is also discussed in Ref.~\cite{Kamiya:2019uiw}. As shown by the dashed line in Fig.~\ref{fig:fit_predict}, an increase in the source size from $R=0.9$ fm to $R=1.6$ fm suppresses the correlation in the low-momentum region $\lesssim 100$ MeV/$c$. This property can be tested by comparing it with experimental analyses performed under different collision conditions and particle species. Indeed, the ALICE data for Pb-Pb collisions~\cite{ALICE:2021szj}, which corresponds to a larger source size of $R\sim 5.2$ fm, experimentally confirms the suppression of the correlation in the low-momentum region. Furthermore, a more quantitative investigation of the source size dependence is conducted through a collaboration of theory and experiment in Ref.~\cite{ALICE:2022yyh}. In this study, theoretical calculation is performed by combining the wave function obtained from the Kyoto potential with the source weights $\omega_{i}$ estimated from the statistical thermal model, and the results are compared with experimental data from $pp$, $p$-Pb, and Pb-Pb collisions, which correspond to various source sizes. A detailed comparison has revealed that reproducing the experimental data requires a larger source weight for the $\bar{K}^{0}n$ channel than the estimated value. This suggests that the coupling to the $\bar{K}^{0}n$ channel may be underestimated, highlighting the potential of high-precision correlation function data to impose new constraints on hadron interactions.

%%%%%%%%%%%%%%%%%%%%
\section{Compositeness of $\Lambda(1405)$}
%%%%%%%%%%%%%%%%%%%%

% intro
Once the basic properties, such as the eigenenergy of the $\Lambda(1405)$ and the $\bar{K}N$ interactions, are precisely determined, the next question is whether the $\Lambda(1405)$ indeed has a hadronic molecular structure. In this section, we introduce the studies on the internal structure of the $\Lambda(1405)$ using compositeness~\cite{Kamiya:2015aea,Kamiya:2016oao}.

% compositeness def
Let us first consider the compositeness of a bound state $\ket{B}$ that exists in a system described by the Hamiltonian $H$. We assume that the Hamiltonian $H$ is decomposed into a free Hamiltonian $H_{0}$ and an interaction term $V$. The complete set of eigenstates of $H_{0}$ consists of a single-particle state $\ket{B_{0}}$ and two-particle scattering states $\ket{\bm{p}}$, labeled by their relative momentum $\bm{p}$. Since $\ket{B_{0}}$ and $\ket{\bm{p}}$ are eigenstates of $H_{0}$, they are orthogonal to each other. The physical bound state $\ket{B}$ can then be expanded as
\begin{align}
   \ket{B} = c\ket{B_{0}}+\int d^{3}p \ \chi(\bm{p})\ket{\bm{p}} .
\end{align}
Here, $\ket{\bm{p}}$ corresponds to a two-particle composite state, and the compositeness $X$ is defined as
\begin{align}
   X = \int d^{3}p
   \ |\braket{\bm{p}|B}|^{2},
\end{align}
which is the overlap integral between the physical state $\ket{B}$ and the scattering states $\ket{\bm{p}}$. This quantity represents the probability of finding the scattering component $\ket{\bm{p}}$ in $\ket{B}$, thus reflecting the fraction of the hadronic molecular component in exotic hadrons.

% weak-binding relation
It should be noted that defining the compositeness $X$ requires the choice of the eigenstates of $H_0$. In other words, the compositeness is dependent on the decomposition of the Hamiltonian as $H = H_0 + V$ and is, therefore, not a strictly observable quantity but generally has model dependence~\cite{Kinugawa:2024crb}. However, there are exceptional cases where the compositeness $X$ can be related to observable quantities. When the state $\ket{B}$ exists near the threshold of $s$-wave scattering, the weak-binding relation holds~\cite{Weinberg:1965zz,Hyodo:2013nka,Kinugawa:2024crb}:
\begin{align} 
   a_0 = R\left\{\frac{2X}{1+X} + \mathcal{O}\left(\tfrac{R_{\rm typ}}{R}\right)\right\},
   \quad R=\frac{1}{\sqrt{2\mu B}},
   \label{eq:weakbinding} 
\end{align} 
where $a_0$, $B$, and $\mu$ are the scattering length, binding energy, and reduced mass, respectively, and $R_{\rm typ}$ represents the typical length scale of the interaction. In Eq.~\eqref{eq:weakbinding}, the model dependence of the compositeness is encapsulated in the higher-order terms $\mathcal{O}(R_{\rm typ}/R)$. Thus, if the binding energy $B$ is sufficiently small near the threshold such that $R \gg R_{\rm typ}$ and the higher-order terms can be neglected, the weak-binding relation allows for the determination of the compositeness in a model-independent manner. This, in turn, enables discussions on the internal structure of hadrons without relying on specific models. 

% generalization to unstable states
Since the original weak-binding relation~\eqref{eq:weakbinding} has an important limitation that it is only applicable to stable bound states, it cannot be directly applied to unstable exotic hadrons such as the $\Lambda(1405)$. The generalization of the weak-binding relation~\eqref{eq:weakbinding} to unstable quasibound states is formulated in Refs.~\cite{Kamiya:2015aea,Kamiya:2016oao}:
\begin{align} 
   a_0 = R\left\{\frac{2X}{1+X} + \mathcal{O}\left(\left|\tfrac{R_{\rm typ}}{R}\right|\right)
   +\mathcal{O}\left(\left|\tfrac{\ell}{R}\right|^{3}\right)\right\},
   \quad R=\frac{1}{\sqrt{-2\mu E_{h}}},
   \label{eq:weakbindingQB} 
\end{align} 
where the third term gives an additional correction, which is characterized by $\ell=1/\sqrt{2\mu \nu}$, the length scale associated with the threshold energy difference $\nu$. Note also that the scattering length $a_{0}$, eigenenergy $E_{h}$, and compositeness $X$ are now all complex numbers, reflecting the unstable nature of the quasibound state. 

% complex compositeness and correction terms
The complex compositeness $X$ poses a challenge to its interpretation. Namely, we need a prescription to define an interpretable fraction of the hadronic molecular component out of the complex $X$~\cite{Aceti:2014ala,Guo:2015daa,Kamiya:2015aea,Kamiya:2016oao,Sekihara:2015gvw,Matuschek:2020gqe,Kinugawa:2024kwb}. In Refs.~\cite{Kamiya:2015aea,Kamiya:2016oao}, the quantity $\tilde{X}$ is proposed as a weight of the molecular component:  
\begin{align} 
   \tilde{X}
   =\frac{1-|1-X|+|X|}{2}.
   \label{eq:Xtilde}
\end{align}  
Furthermore, based on the correction terms in Eq.~\eqref{eq:weakbindingQB}, a numerical method is developed to evaluate the uncertainty of compositeness. 

\begin{figure}[bt]
\centering
\includegraphics[width=8cm]{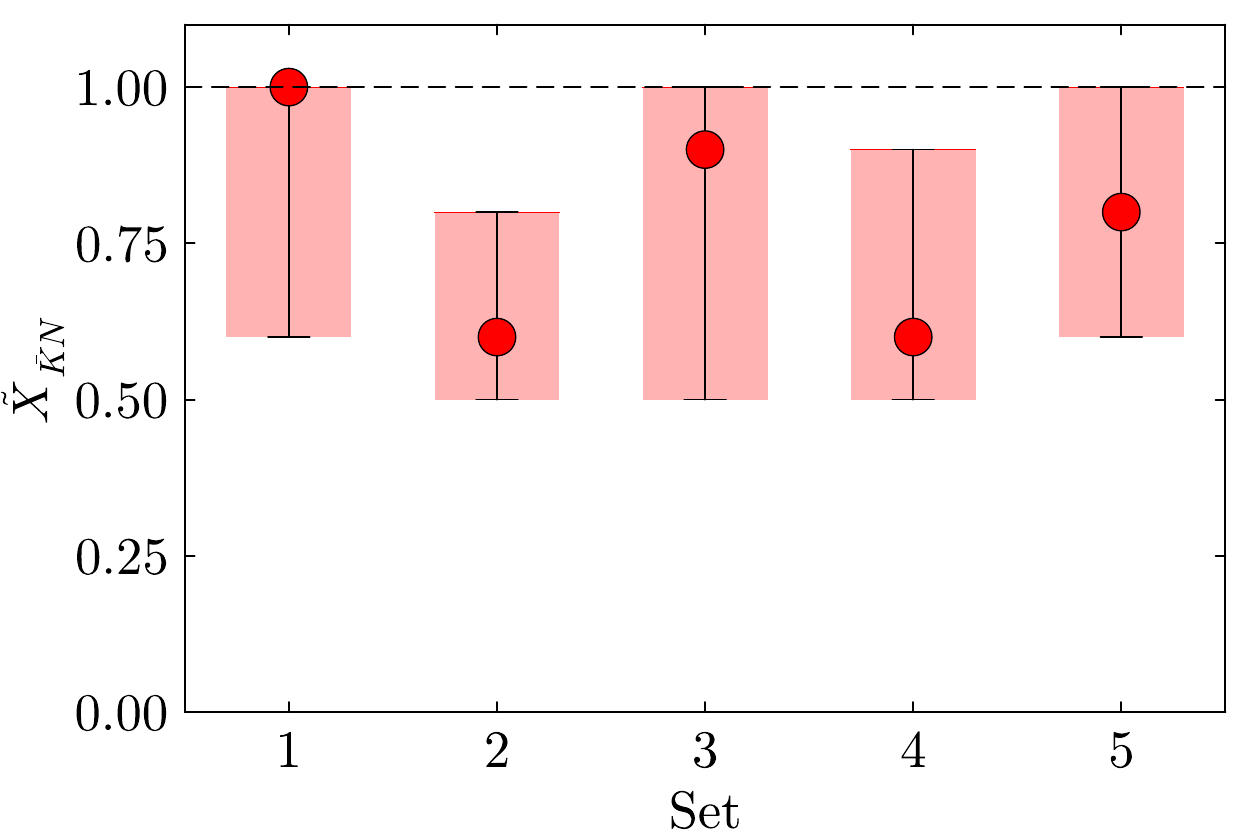}
\caption{$\bar{K}N$ compositeness $\tilde{X}_{\bar{K}N}$ of the $\Lambda(1405)$ with uncertainty estimation~\cite{Kamiya:2015aea,Kamiya:2016oao}.}
\label{fig:1405compositeness}
\end{figure}

% application to Lambda(1405)
Now we are in a position to evaluate the compositeness of actual exotic hadrons. Based on the results of the analysis of chiral coupled-channel approaches~\cite{Ikeda:2011pi,Ikeda:2012au,Mai:2012dt,Guo:2012vv,Mai:2014xna}, we prepare sets of the $\bar{K}N$ scattering length $a_{0}$ and eigenenergy $E_{h}$ of the $\Lambda(1405)$, labeled as Set 1 to Set 5. For each set, we apply Eq.~\eqref{eq:weakbindingQB} while neglecting the correction term to obtain the complex $\bar{K}N$ compositeness $X_{\bar{K}N}$. By substituting this into Eq.~\eqref{eq:Xtilde}, we determine the central value of the physically interpretable compositeness $\tilde{X}_{\bar{K}N}$. Furthermore, estimating the $\bar{K}N$ interaction range from $\rho$ meson exchange as $R_{\rm typ} \sim 0.25$ fm and using $\ell \sim 1.08$ fm, determined from the energy difference to the $\pi\Sigma$ threshold, we evaluate the correction terms. The results shown in Fig.~\ref{fig:1405compositeness} demonstrate that $\tilde{X}_{\bar{K}N}$ remains greater than 0.5 even when including the uncertainty band, providing a quantitative indication that the internal structure of $\Lambda(1405)$ is predominantly composed of the $\bar{K}N$ molecular component.  

%%%%%%%%%%%%%%%%%%%%
\section{Summary and future prospects}
%%%%%%%%%%%%%%%%%%%%

% summary
In this contribution, we review recent studies on hadron interactions, which are essential for investigating the hadronic molecular components of exotic hadrons. We show that the femtoscopy technique using momentum correlation functions has emerged as a novel method for studying hadron interactions. In particular, precise measurements of the $K^-p$ correlation functions provide crucial insights into the $\Lambda(1405)$ and $\bar{K}N$ interactions. Furthermore, the evaluation of the compositeness of the $\Lambda(1405)$ based on the weak-binding relation indicates the dominance of the $\bar{K}N$ molecular component, utilizing observable quantities such as the scattering length and eigenenergy.

% future prospects
The study of hadron interactions is expected to continue advancing through the synergy of modern scattering experiments at J-PARC, femtoscopy, and lattice QCD calculations. In particular, the precise determination of scattering lengths and eigenenergies plays a crucial role in investigating hadron internal structures through compositeness. One of the new directions being explored at J-PARC is the femtoscopy of correlation functions involving light nuclei, which may be realized in the J-PARC Heavy-Ion (HI) program. In fact, new insights into hyperon interactions could be obtained from correlation functions such as $\Lambda\alpha$ and $\Xi\alpha$~\cite{Jinno:2024tjh,Kamiya:2024diw}, and further developments in this field are highly anticipated.

%%%%%%%%%%%%%%%%%%%%
\section*{Acknowledgments}
%%%%%%%%%%%%%%%%%%%%

The author is grateful to Yudai Ichikawa and Kyoichiro Ozawa for the invitation to this symposium. Appreciation is also extended to Yuki Kamiya, a collaborator on the main results of this contribution. This work has been supported in part by the Grants-in-Aid for Scientific Research from JSPS (Grants
No.~JP23H05439 and % Kiban S (Hyodo)
No. JP22K03637). % Kiban C (Hyodo)

\end{document}